\documentclass[journal=jacsat,manuscript=article]{achemso}

\usepackage[version=3]{mhchem} 

\title{Implantation and atomic scale investigation of self-interstitials in graphene}

\author{Ossi Lehtinen}
\email{ossi.lehtinen@gmail.com}
\author{Nilesh Vats}
\author{Gerardo Algara-Siller}
\author{Pia Knyrim}
\author{Ute Kaiser}
\affiliation[Ulm University]{Central Facility for Electron Microscopy, Group of Electron Microscopy of Materials Science, Ulm University, Germany}

\keywords{graphene, implantation, self-interstitials, defects, HRTEM, MD}

\begin{document}

\begin{abstract}
Crystallographic defects play a key role in determining the properties of crystalline materials. The new class of two-dimensional materials, foremost graphene, have enabled atomically resolved studies of defects, such as vacancies~\cite{meyer2012,kotakoski2011,kotakoski2011b,standop2013}, grain boundaries~\cite{yazyev2010,huang2011,kurasch2012}, dislocations~\cite{warner2012,lehtinen2013}, and foreign atom substitutions~\cite{cretu2010,meyer2011,zhou2012,bangert2013,robertson2013}. However, atomic resolution imaging of implanted self-interstitials has so far not been reported in any three- but also not in any two-dimensional material. Here, we deposit extra carbon into single-layer graphene at soft landing energies of $\sim$1~eV using a standard carbon coater. We identify all the self-interstitial dimer structures theoretically predicted earlier~\cite{orlikowski1999,adatoms1,istw}, employing 80~kV aberration-corrected high-resolution transmission electron microscopy. We demonstrate accumulation of the interstitials into larger aggregates and dislocation dipoles, which we predict to have strong local curvature by atomistic modeling, and to be energetically favourable configurations as compared to isolated interstitial dimers. Our results contribute to the basic knowledge on crystallographic defects, and lay out a pathway into engineering the properties of graphene by pushing the crystal into a state of metastable supersaturation.
\end{abstract}

There is no such thing in nature as a perfect crystal, although many materials that are useful in engineering are crystalline to a good approximation, with the exception of few volume defects and surfaces interrupting the perfect periodicity. The defects, however, play a key role in determining the properties of materials. Therefore an important branch in materials science is dedicated to studying, understanding and controlling defects in the crystal volume in order to understand and tailor the properties of materials. For example, control over the electronic structure of semiconductors is achieved through introduction of zero-dimensional defects (impurity atoms) into the crystals, the mechanical properties of metals and alloys are to a large extent controlled by their one-dimensional (dislocations) and two-dimensional (grain boundaries) defects, and the properties of the two-dimensional surfaces of materials are altered by surface reconstructions and can be manipulated by bonding of foreign atom species on the surface. 

The introduction of graphene and other two-dimensional (2D) crystals into the zoo of known materials has opened up a completely new perspective into studies of crystallographic defects. This is, most of all, due to the simple fact that in 2D materials the three-dimensional bulk of a crystal does not obscure the view of the defects, as surface is essentially all that the materials have. With modern microscopy methods, such as aberration-corrected high-resolution transmission electron microscopy (AC-HRTEM), aberration-corrected scanning transmission electron microscopy and scanning tunneling microscopy, the exact atomic structure of the 2D materials can be resolved~\cite{temgeneral,krivanek2010,ndiaye2008}, and crystallographic defects, and their dynamics can be studied at the level of the basic building blocks of matter~\cite{defectdynamicsgeneral}.

A large body of work has been dedicated to studying point defects and extended defects in graphene. Vacancies, introduced for example by ion~\cite{standop2013} or electron irradiation~\cite{meyer2012} have been observed, along with their transformations~\cite{kotakoski2011}, migration, and coalescence~\cite{kotakoski2011b}. Foreign atoms at substitutional sites or adsorbed to the graphene lattice have been detected and identified~\cite{cretu2010,meyer2011,zhou2012,bangert2013,robertson2013}. The structure and movement of grain boundaries have been observed~\cite{yazyev2010,huang2011,kurasch2012}, and studies of dislocations in graphene have allowed resolving their exact atomic structure~\cite{warner2012,butz2013dislocations} and provided the first real-time atomic-scale observations of the full life-cycle of a dislocation --- a long standing topic in materials science~\cite{lehtinen2013}. 

\begin{figure*}
	\includegraphics[width=.95\textwidth]{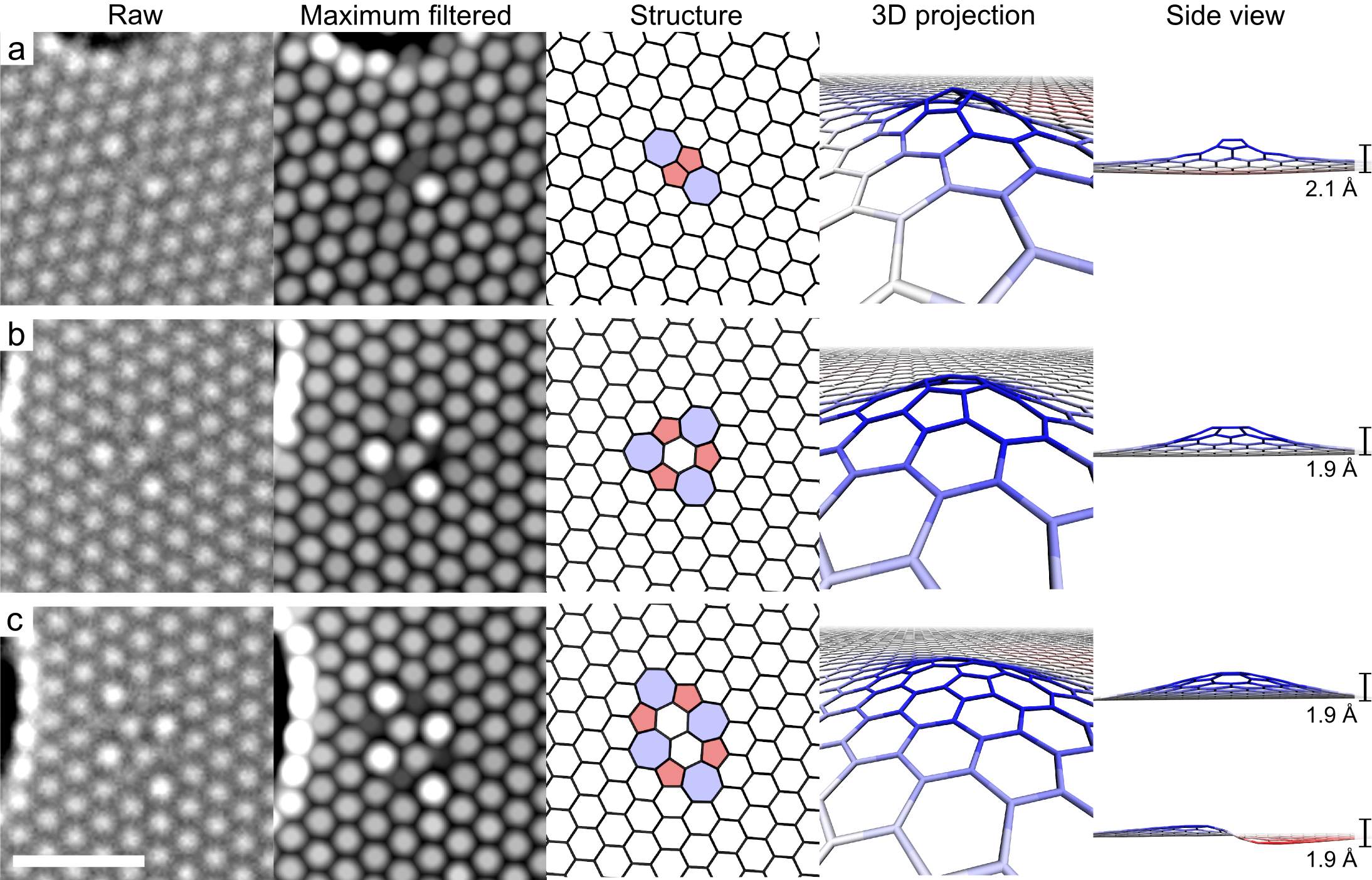}
	\caption{{\bf AC-HRTEM characterization and structural modeling of carbon deposited on graphene.} {\bf a}: The inverse Stone-Thrower-Wales defect, {\em i.e.} a self-interstitial dimer. {\bf b}: A self-intestitial dimer after a single bond-rotation. {\bf c}: A self-interstitial dimer after a second bond-rotation. The first column shows the raw HRTEM images with corrected three-fold astigmatism as explained in the Supplementary figure 3 and Ref.~\cite{lehtinen2014numerical}, the second column show the HRTEM images after maximum filtering, which improves the visibility of the structure~\cite{lehtinen2013}, the third column shows structural models of the defects (pentagonal carbon rings are colored red and heptagons blue), the fourth column shows 3D projections of the relaxed structures, and the fifth column shows a side view of the relaxed structures, displaying strong out of plane buckling. The third structure relaxed in symmeteric and anti-symmetric modes, as shown in the side view. The scale bar is 1~nm.}
	\label{fig1}
\end{figure*}

The bonding sites of single carbon adatoms~\cite{singleadds1,singleadds2,ma2007}, and stable configurations of fully sp$^2$-coordinated interstitial dimers incorporated in the graphene lattice~\cite{orlikowski1999,adatoms1,istw} (see ~\ref{fig1} for the atomic structures) have been theoretically predicted. Experimental evidence of single carbon adatoms in the so-called bridge position (with the extra atom sitting on top of a C-C bond of graphene) has been presented~\cite{meyer2008b}. Structures analogous to the interstitial dimers in graphene have been observed in incompletely crystallized hexagonal 2D silicon oxide monolayers using both scanning tunneling microscopy~\cite{yang2013} and AC-HRTEM~\cite{bjorkman2013}. According to calculations, the extra atom structures are expected to modify the electronic and magnetic properties of graphene~\cite{singleadds1,istw}, and increase the chemical reactivity locally~\cite{istw}, which becomes important when functionalizing graphene. The interstitial dimer structures have been observed in earlier studies~\cite{lehtinen2013,bjorkman2013,robertson2014} after exposing single-layer graphene to extreme electron doses in a TEM, but importantly, these structures did not result from introduction of extra atoms in the crystal, but rather from removal of a large number of atoms around the defects and significant reordering of the whole graphene lattice. Furthermore, with the electron irradiation approach, the interstitial defects are randomly appearing among a wide variety of other types of defects, and one has essentially no control over which defects are introduced into the graphene lattice. Importantly, observation of isolated self-interstitial dimers in non-treated graphene samples has not been reported, despite extensive atomic scale studies of both mechanically exfoliated and chemical vapour deposition (CVD) grown graphene.

In terms of deviations from the equilibrium density of the graphene crystal, deficit type defects, such as single vacancies and vacancy agglomerates, can be introduced by removing atoms from the lattice. In a TEM this can be accomplished by simply exposing graphene to the electron beam of the microscope, as a single knock-on collision event between an electron and a carbon atom in graphene is adequate for removing the atom at voltages of 80~kV and above, leaving a vacancy behind~\cite{meyer2012}. On the other hand, an external source of new carbon atoms needs to be available in order to produce density surplus type of defects. As the TEM is an ubiquitous tool for characterizing defects in graphene and other 2D-materials, it is understandable that observations of deficit type defects are abundant, where as reports on density surpluses are scarce. What further complicates studies of surplus atoms by TEM, is the fact that the extra atoms themselves can be knocked out by the electron beam. As the displacement threshold energy of atoms at defect sites tends to be lower than in pristine graphene~\cite{lehtinen2013,susi2012}, special care needs to be taken in order to minimize the electron dose on the samples.

In general, it is energetically expensive to introduce a positive density deviation into any crystal due to the large required strain. However, graphene behaves in a special way in the presence of extra atoms, as the graphene plane can deform in the third out-of-plane direction, and due to the small bending modulus of graphene~\cite{yakobson1996}, the energy cost of adding extra atoms in graphene is relatively low. Interestingly, especially in the cases of the reconstructed interstitial dimers (see \ref{fig1} b and c for the atomic structures), the two extra atoms are incorporated in the lattice in such a way that it is impossible to pinpoint which exact atoms are the surplus ones. An intriguing alternative interpretation of the reconstructed interstitial dimer structures is to view them as miniscule grain boundary loops with an associated density surplus~\cite{cockayne2011}, since a chain of pentagons and heptagons are enclosing one or two carbon hexagons in the defects. The line between adatoms and interstitials becomes fuzzy in the case of 2D materials such as graphene, and it is debatable into which category each extra atom structure belongs to. Here, we elect to term single atoms residing on top of the graphene lattice as adatoms, and atoms incorporated into the lattice through perfect sp$^2$-bonding as self-interstitials.

In this Letter, we report on implantation of extra carbon into single-layer graphene at soft landing energies in the range of 1~eV. The resulting structures and their electron irradiation induced dynamics are characterized by AC-HRTEM operated at 80~kV. All the three theoretically predicted self-interstitial dimer structures~\cite{orlikowski1999,adatoms1,istw} are observed in the samples. Furthermore, larger aggregates of sp$^2$-bonded extra atoms, as well as dislocation dipoles with the associated local density surplus are observed. With the help of atomistic simulations we show that such defect sites have high local curvature, leading to blister-like structures. After extended electron irradiation, the extra atoms are removed from the crystal, leaving behind a pristine graphene lattice. Altogether, our results present the first atom-by-atom resolved study of a crystal forced into a state of supersaturation, that is, containing self-interstitials. Further on, our results lay out a pathway into engineering the properties of graphene by implantation of self-interstitials into the lattice.

The method for implanting carbon into graphene was straightforward: First, mechanically exfoliated and commerically obtained CVD graphene was transferred onto Quantifoil TEM grids. The samples were then treated in a carbon coating apparatus normally used for depositing a conductive amorphous carbon film on top of non-conducting specimens for electron microscopy. In this apparatus, the sample and a graphitic filament are placed in a vacuum chamber, and a current is run through the filament in order to bring it to its sublimation temperature. The system was run with parametrization close to minimum achievable deposition thickness (one to four 100-300~ms pulses). In addition to indivudual C atoms, also molecules such as C$_2$ and C$_3$ are produced when sublimating graphite~\cite{drowart1959}, and all these land on the sample surface (and other surfaces in the chamber) at thermal kinetic energies, typically less than 1~eV/atom. 

We conducted analytical potential molecular dynamics simulations for predicting the sticking probability of landing C atoms and C$_2$ dimers at different kinetic energies, and based on these results we expect a large fraction of the incoming atoms/molecules to form bonds with graphene, when the kinetic energy is less than 30~eV/atom (see Supplementary figure 1 for sticking probabilities as a function of kinetick energy). Earlier theoretical predictions on B and N implantation~\cite{ahlgren2011} also suggest the possibility of implanting atoms into graphene, and in a recent experiment B and N atoms were, in fact, implanted in free-standing graphene~\cite{bangert2013} at a landing energy of 20~eV. Individual carbon adatoms are predicted to be mobile at room temperature~\cite{singleadds1,ma2007}, and are thus not expected to be found after deposition, although, upon encountering another carbon adatom, a highly stable self-interstitial dimer can be formed~\cite{adatoms1,istw}. The mobility of the larger molecules can be expected to be lower, but no theoretical predicions on such mobilities could be found, and conducting such simulations is beyond the scope of this study. Spatial control over the landing site distribution cannot be achieved using our approach, and more sophisticated methods are required if precise engineering of graphene by introduction of interstitials is desired. 

The samples were characterized using AC-HRTEM after the carbon deposition treatment. Varying coverage of the free-standing graphene layer was observed depending on the deposition parametrization, proving that at least a fraction of the landing atoms/molecules are sticking to the sample surface. When the carbon coater was run at low deposition rates ({\em e.g.}, one 200 ms pulse) isolated dark spots were frequently observed. Based on contrast analysis, the dark spots could be interpreted as individual carbon adatoms (see Supplementary figure 2 for details). However, their remarkable stability is inconsistent with theoretical predictions on the migration barriers~\cite{singleadds1,ma2007} suggesting high mobilities at room temperature. Thus, we conclude that the spots are likely other types of defects, such as adsorbed molecules like CH$_3$~\cite{erni2010} or even silicon in a substitutional position~\cite{zhou2012}.

When it comes to self-interstitial dimers, the situation is completely different in terms of identifiability of the defects. The interstitial dimers are predicted to be incorporated into the sp$^2$-coordinated graphene lattice with no dangling bonds, resulting in high stability of the defects~\cite{istw,adatoms1}. Distinct polygon patterns formed out of pentagons, hexagons, and heptagons can be used to unambiguously identify these point defects (see \ref{fig1} as well as Supplementary figure 3 for discussion on the effects of residual A2 astigmatism). In fact, the so-called inverse Stone-Thrower-Wales defect (\ref{fig1} a)~\cite{istw}, and both its other two polymorphs (resulting from rotation of one and two C-C bonds)~\cite{orlikowski1999,adatoms1} were observed in the samples (\ref{fig1} b and c). Thus, the theoretical predictions of the structure of these defects in isolated form are experimentally verified, and the first atomic resolution images of implanted self-interstitials in any material at atomic resolution are presented.

The formation process of the self-interstitial dimers cannot be captured in the microscope, as all the observed defects are in place when the first HRTEM images are obtained. The electron beam of the microscope may, indeed, play a role in the formation of the final stable defects from possibly less ordered structures after implantation. The electrons provide energy to the system, and thus can allow the atoms to overcome activation barriers, similar to what has been observed, for example, in the case of electron beam stimulated bond-rotations~\cite{kotakoski2011} and grain-boundary migration~\cite{kurasch2012} in graphene. Inverse Stone-Thrower-Wales defects have been predicted to form most readily next to divacancy defects in graphene~\cite{adatoms1}. However, vacancies next to any of the self-interstitial structures were never observed in our experiment. As demonstrated below, in addition to possibly helping in formation of the observed defects, the impacts of the energetic electrons can also knock out the extra atoms into the vacuum of the microscope~\cite{lehtinen2013,susi2012}, thus eliminating the defects. 

\begin{figure*}
	\includegraphics[width=.95\textwidth]{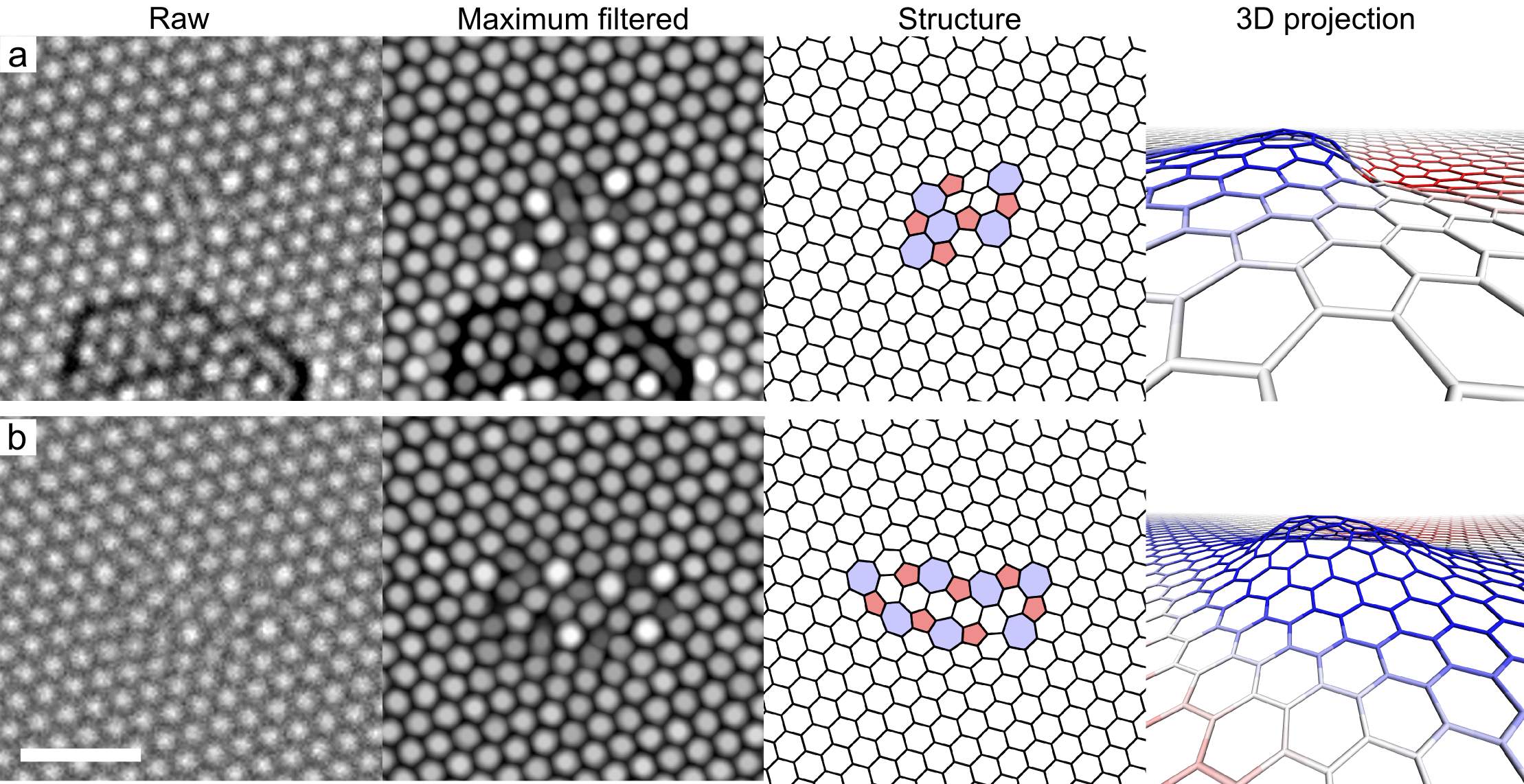}
	\caption{{\bf Carbon interstitial aggregates in graphene.} {\bf a}: An interstitial aggregate consisting of four extra carbon atoms. {\bf b}:  An interstitial aggregate with ten extra carbon atoms. The first column shows the raw HRTEM images, the second column show the HRTEM images after maximum filtering, which improves the visibility of the structure~\cite{lehtinen2013}, the third column shows structural models of the defects (pentagonal carbon rings are colored red and heptagons blue), and the fourth column shows 3D projections of the relaxed structures. The scale bar is 1~nm.}
	\label{fig2}
\end{figure*}

The atomic models of the interstitial defects shown in \ref{fig1} and \ref{fig2} were relaxed using the conjugate gradient algorithm, with interatomic forces described by a valence force field model (VFF)~\cite{vff}. In each case strong out-of-plane buckling was observed (\ref{fig1}), extending over long distances, and in order to accommodate this strain field in the simulation cell, a system consisting of 20002 atoms was used in the simulations. The relaxed structures extend 1.9-2.1~{\AA} out of the plane. The calculated formation energies of the defects in panels a, b, and c are 5.8~eV, 6.2~eV, and 7.1~eV, respectively. The third defect relaxed in two different modes, extending symmetrically or antisymmetrically out of the plane, with the antisymmetric mode resulting in only a slightly higher formation energy of 7.2~eV than the symmetric mode (7.1~eV). The formation energies of these defects have been calculated earlier using density functional theory (DFT)~\cite{adatoms1,istw}. The values disagree by tenths of eVs, which can be explained by the higher level of approximation in our method, as well as the unavoidable limited system sizes in the DFT calculations. In this work we opt to use the VFF model as it allows us to model the long ranging corrugations connected to the defects.

The interstitial dimers were observed to undergo transformations under the electron beam. \ref{fig1} b and c show the same interstitial dimer before and after a bond-rotation event. Such transformations back and forth between these two configurations were frequently observed, but curiously transformation to and from the inverse Stone-Thrower-Wales configuration (\ref{fig1} a) was never observed, even though the predicted formation energy of this configuration is lower than the others'. Further on, the lowest energy configuration was observed most seldom of the three. This, however, might be the result of a selection bias, as it is less visible in the micrographs as compared to the others. All of the interstitial dimers eventually disappeared due to knocking out of the extra atoms by the electron beam after doses in the order of 10$^9$ electrons/nm$^2$.

Larger aggregates of interstitials in the lattice were observed as well (see \ref{fig2}). The two cases presented in panels a and b of \ref{fig2} show structures with four and ten extra atoms, respectively. The relaxed atomic models of the defects display again strong out of plane buckling. The calculated formation energies of the four extra atom and ten extra atom structures are 11.4~eV and 14.8~eV, respectively. Normalizing to energy per interstitial dimer (5.7~eV and 3.0~eV), one can make a comparison to the formation energies of the isolated interstitial dimers, and conclude that it is energetically favourable for the interstitials to agglomerate. It should be pointed out, that these are not necessarily the optimal configurations for each number of interstitials, but rather cases which were experimentally observed.

In many cases contamination was seen to stick to the interstitial defects. The contamination was removed by the electron beam after prolonged exposure, revealing the sp$^2$-coordinated defect structures. The sticking of contamination indicates a higher affinity of molecules adsorbed on top of graphene to the interstitials, which in turn is promising in terms of functionalization of graphene, {\em e.g.}, when graphene is used as a sample support for studies of nano objects in a TEM~\cite{Pantelic2012}.

\begin{figure}
	\includegraphics[width=.5\textwidth]{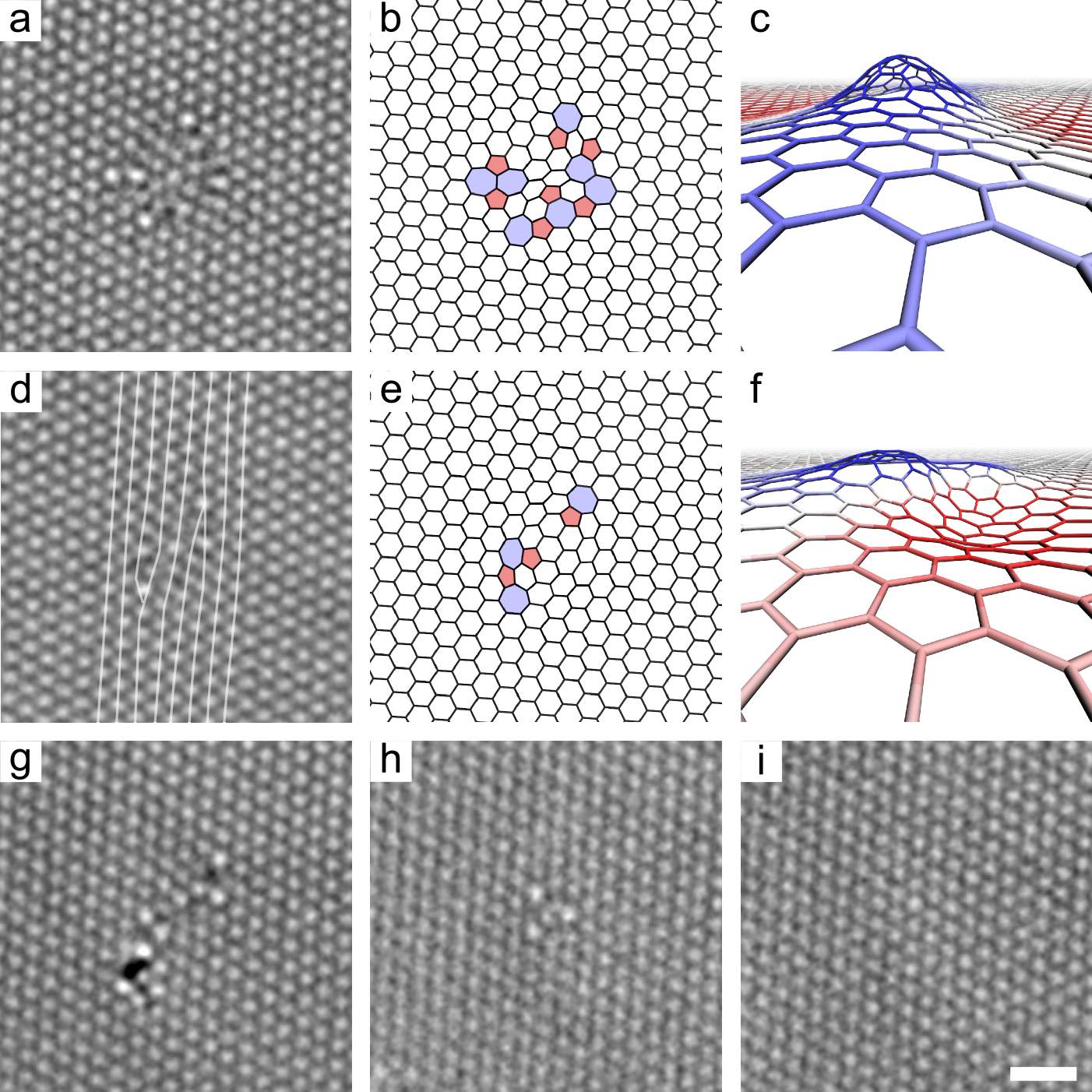}
	\caption{{\bf Evolution of an interstitial aggregate to a dislocation dipole and eventual disappearance.} {\bf a}: An interstitial aggregate. {\bf b}: Its approximate structural model. {\bf c}: A 3D projection of the relaxed structure. {\bf d}: The same defect after an electron dose of $1.4\times10^9$ e/nm$^2$. The extra carbon atoms have rearranged into a dislocation dipole. {\bf e}: A structural model of the previous. {\bf f}: A 3D projection of the previous. {\bf g}, {\bf h}, and {\bf i}: Further evolution of the structure, with an interstitial dimer preceding the complete disappearance of the defect, after a total electron dose of $6\times10^9$ e/nm$^2$. The scale bar is 1~nm.}
	\label{fig3}
\end{figure}

Similar to the earlier reported case of vacancy aggregates~\cite{vff,lehtinen2013}, there is a possibility for the graphene lattice to reorganize into a dislocation dipole in the presence of extra atoms. In such a situation, the extra atoms occupy an extra row in between the dislocation cores. Such rearrangements were observed in our experiment as presented in \ref{fig3}. First, a rather disordered aggregate of extra atoms was observed (panel a). Here, the structural model should be considered as an approximate interpretation of the defect, as the structure was constantly changing during imaging. Additionally, the dark high-contrast spots in the frame (and some of the subsequent frames) suggest the presence of, {\em e.g.}, sp$^3$-coordinated C atoms, which make the exact interpretation based on the projected TEM-image difficult. Nevertheless, even if such atoms are neglected in the model, a density surplus is identified.

Under continuous electron irradiation, the atoms rearrange into a dislocation dipole, as can be seen in \ref{fig3} d. The distance of the dislocation cores is five lattice rows, corresponding to ten extra carbon atoms in the lattice. The lower dislocation core deviates from the simple pentagon hexagon structure due to a single rotated bond. The calculated formation energy of this structure is 10.82~eV, or 2.16~eV per interstitial dimer. In the subsequent frames g, h, and i the defect is observed to further transform, with gradual removal of surplus atoms, and shrink into an interstitial dimer in the second to last configuration (panel h). Eventually the defect disappears, leaving pristine graphene behind.

\begin{figure}
	\includegraphics[width=.5\textwidth]{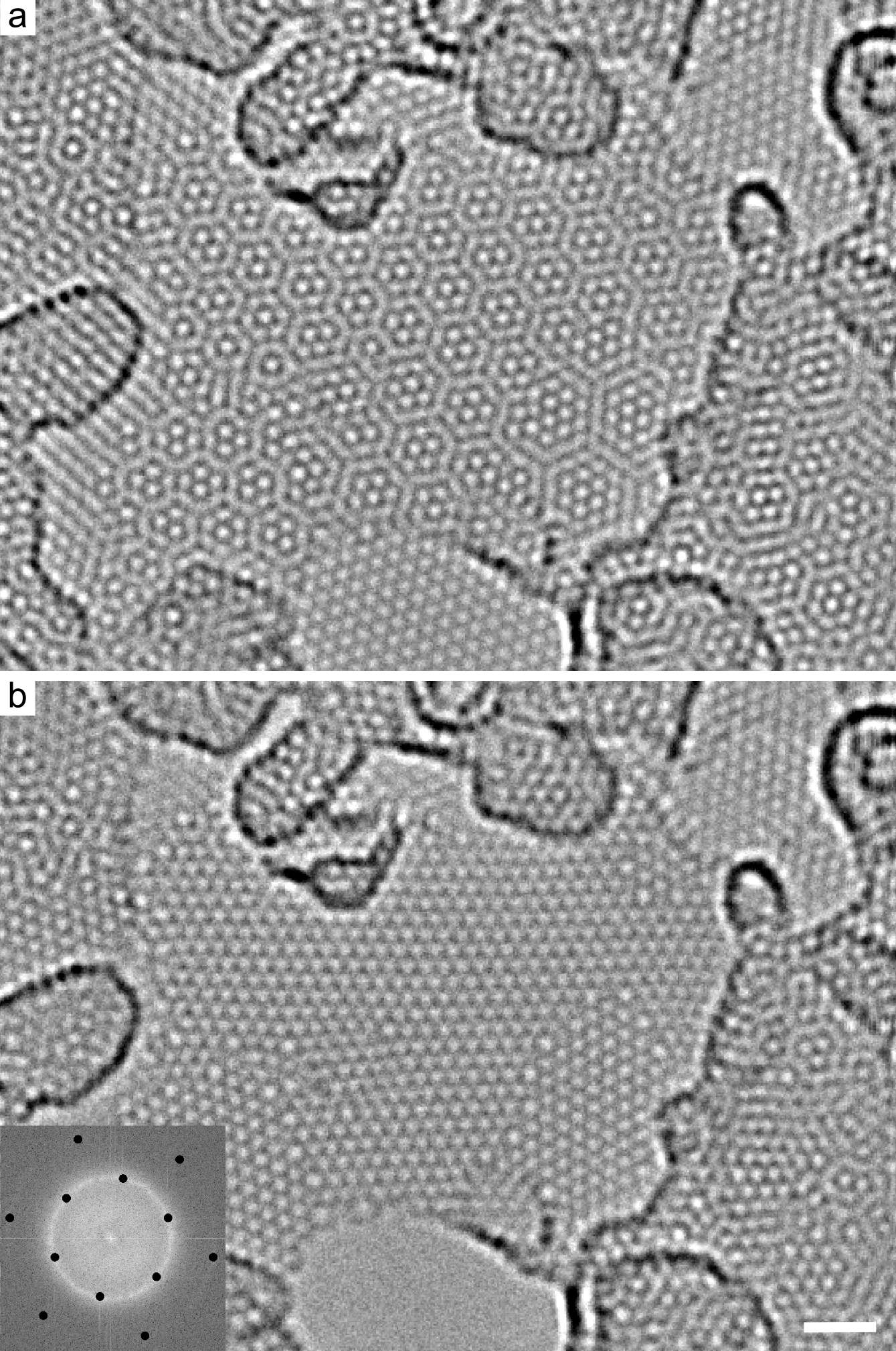}
	\caption{{\bf Second nanocrystalline graphene layer on top of graphene.} {\bf a}: An AC-HRTEM image of the second layer appearing as regular Moire patterns in the micrograph. {\bf b}: The same frame after the image contribution of first graphene layer is removed by Fourier filtering~\cite{fourier}. The inset of b shows the mask used for the filtering. The scale bar is 1~nm.}
	\label{fig4}
\end{figure}

A limit in the density of interstitials in the graphene layer that can be introduced is reached if the deposition thickness is increased to approximately one monolayer. Instead of atoms incorporated in the graphene lattice, the extra carbon is segregated into a new layer. Interestingly, high-magnification images of the second layer showed the formation of a nanocrystalline graphene layer on top of graphene (\ref{fig4} a). When the image contribution of the first graphene layer is removed by Fourier filtering~\cite{fourier} the few nanometer sized graphene grains become visible, separated by continuous chains of pentagonal and hexagonal carbon rings (\ref{fig4} b). The new layer was rather non-uniform, but the uniformity was observed to improve during imaging, facilitated by the impacts of the electrons allowing the system to overcome configuration barriers, analogous to high temperature annealing~\cite{kurasch2012}.

To conclude, we have introduced a surplus density of carbon into suspended graphene by means of low-energy implantation of carbon atoms resulting ultimately in self-interstitial dimers incorporated in the graphene crystal. The implantation was conducted using an evaporating carbon coating apparatus, which works by bringing a graphitic filament to its sublimating temperature, and thus emitting individual C atoms, C$_2$, C$_3$ and larger molecules which then land on the sample surface at thermal kinetic energies. By careful tuning of the deposition parameters, a low enough density of extra carbon was reached in order to produce isolated point defects. However, when the deposition thickness was increased to approximately one carbon monolayer, instead of a high density of interstitials in the original graphene layer, a second nanocrystalline graphene layer was formed. The structure and electron irradiation induced dynamics of the produced defects was imaged at the atomic scale, employing 80~kV aberration-corrected high-resolution transmission electron microscopy. All the earlier theoretically predicted, completely sp$^2$-coordinated structural configurations of isolated self-interstitial dimers in graphene were experimentally verified. Additionally, larger aggregates of interstitials and edge dislocation dipoles incorporated in the graphene lattice were observed, and based on atomistic modeling, such structures were determined to be energetically favourable arrangements for the extra atoms. All of the interstitial structures were predicted to strongly buckle out-of-plane. Such blister-like structures can be expected to have higher reactivity than pristine graphene, which can be advantageous for functionalization of graphene. Further on, defect structures containing surplus carbon atoms have been predicted to have exciting electronic and magnetic properties~\cite{singleadds1,istw}, and our experiments demonstrate that such structures can, in fact, be fabricated.

\section{Methods}

\subsection{Experimental} The graphene samples were produced either via mechanical exfoliation of natural graphite, or from commercially obtained CVD-graphene (Graphenea S.A.), using the methods presented in Refs.~\cite{meyer2008c} and~\cite{Pantelic2011a} for transfering the graphene layer onto TEM-grids.

The carbon deposition was conducted using a Quorum Technologies K950X Turbo Evaporator carbon coater. During normal operation, parameters resulting in an amorphous carbon layer of few nanometers are used, but by running the system at parametrization leading to minimal deposition thickness (one to four 100-300~ms pulses) submonolayer to monolayer deposition thickness could be achieved. The deposition was conducted by passing a current through a graphitic filament, which resulted in thermal sublimation of carbon atoms from the filament. The ejected carbon atoms landed on the suspended graphene samples placed $\sim$~5~cm away from the filament. Energy dispersive X-ray spectroscopy was used to estimate the purity of the graphitic filament in a Zeiss NVision 40 dual-beam FIB/SEM system. The filament was found to contain 97$\pm$2 \% carbon in agreement with the manufacturer's specification.

The atomic scale characterization was conducted using an FEI Titan 80-300 with post specimen hardware spherical aberration correction operated at a voltage of 80~kV. The spherical aberration was corrected down to $\sim$20~$\mu$m, and the extraction voltage of the field emission gun was set to 2~kV in order to reduce the energy spread of the beam. The imaging was done at underfocus conditions, leading to dark atom contast. The electron dose rates were in the range of 2$\times10^7$ e/nm$^2$/s.

\subsection{Computational} The structural relaxations were conducted using the conjugate gradient algorithm, with interatomic forces described by a valence force field model~\cite{vff}, which is fitted to reproduce formation energies of fully sp$^2$-coordinated defects in graphene accurately when compared to density functional theory calculations.

The HRTEM image simulations for the Supplementary figures 2 and 3 were conducted using the QSTEM software package~\cite{qstem}. In the simulations, the spherical aberration was set to 20~$\mu$m, focal spread to 9~nm, and the A2 astigmatism to 0--100~nm in order to reproduce the slighlty higher contrast of the second graphene sublattice in the experimental micrographs.

In the C atom and C$_2$ dimer deposition simulations the interatomic forces were described by an analytical force field~\cite{brenner}, which is computationally efficient enough for simulating a large number of events required for gathering sufficient statistics. In the simulations, impacts of individual carbon atoms and dimers landing in the normal direction of the graphene target consisting of 800 atoms were simulated. The considered kinetic energies ranged from 0.05 to 100~eV per atom with four energy values per order of magnitude. 100 impact simulations per energy at randomized points in the graphene unit cell were run. The orientation of the dimers was randomized, but the role of rotational degrees of freedom was not explored.

\acknowledgement

The authors acknowledge financial support by the DFG (German Research
Foundation) and the Ministry of Science, Research and the Arts (MWK) of Baden-
Wuerttemberg in the frame of the SALVE (Sub Angstrom Low-Voltage Electron
microscopy) project and the financial support by the DFG SPP 1459 'Graphene' project. O.L. acknowledges financial support by the Finnish Cultural Foundation.

{\bf Author contributions} O.L. designed the experiment. N.V. ang G.A.S. prepared the graphene samples. N.V. conducted the carbon deposition. P.K. and O.L. implemented the valence force model code and N.V. and O.L. conducted the atomistic simulations. O.L. conducted the HRTEM imaging, HRTEM image simulations, and prepared the manuscript with assistance from all the authors. N.V. and O.L. analyzed the HRTEM images. U.K. assembled the team and supervised the project.

\suppinfo

Sticking probability for C atoms and C$_2$ dimers landing on graphene. Contrast analysis of isolated atoms on graphene after implantation. Effects of residual three-fold astigmatism in the HR-TEM images. This material is available free of charge via the Internet at \url{http://pubs.acs.org}.


\providecommand*{\mcitethebibliography}{\thebibliography}
\csname @ifundefined\endcsname{endmcitethebibliography}
{\let\endmcitethebibliography\endthebibliography}{}

\end{document}